%% file: Template.tex
\documentclass{article}
\usepackage{spconf,amsmath,graphicx}
\usepackage{multirow}
\usepackage{hyperref}
\usepackage{graphicx}
\usepackage{subcaption}

\title{Mitigating Intra-Speaker Variability in Diarization\\with Style-Controllable Speech Augmentation}
\name{
\begin{tabular}{c}
Miseul Kim$^1$\quad Soo Jin Park$^2$\quad Kyungguen Byun$^2$ \\
\textit{Hyeon-Kyeong Shin$^2$}\quad \textit{Sunkuk Moon$^2$}\quad \textit{Shuhua Zhang$^2$}\quad \textit{Erik Visser$^2$}
\end{tabular}
}

\address{$^1${Department of Electrical and Electronic Engineering, Yonsei University, Seoul, South Korea}\\
$^2${Qualcomm Technologies, Inc., San Diego, California, USA}}

\begin{document}
%\ninept
%
\maketitle
\renewcommand{\thefootnote}{}
\footnotetext{This work was completed as part of an internship at Qualcomm.}
\renewcommand{\thefootnote}{\arabic{footnote}}

\input{contents/0_abstract}
\input{contents/1_intro}
\input{contents/2_proposed}
\input{contents/3_exp_setup}
\input{contents/4_exp}
\input{contents/5_conclusion}

\clearpage
\bibliographystyle{IEEEtran}
\bibliography{refs}

\end{document}

%% file: contents/0_abstract.tex
\begin{abstract}
Speaker diarization systems often struggle with high intrinsic intra-speaker variability, such as shifts in emotion, health, or content. 
This can cause segments from the same speaker to be misclassified as different individuals, for example, when one raises their voice or speaks faster during conversation. 
To address this, we propose a style-controllable speech generation model that augments speech across diverse styles while preserving the target speaker’s identity. 
The proposed system starts with diarized segments from a conventional diarizer. 
For each diarized segment, it generates augmented speech samples enriched with phonetic and stylistic diversity.
And then, speaker embeddings from both the original and generated audio are blended to enhance the system’s robustness in grouping segments with high intrinsic intra-speaker variability.
We validate our approach on a simulated emotional speech dataset and the truncated AMI dataset, demonstrating significant improvements, with error rate reductions of 49\% and 35\% on each dataset, respectively.

\end{abstract}

%% file: contents/1_intro.tex
\section{Introduction}
\label{sec:intro}
Speaker diarization refers to the task of determining “who speaks when” in an audio recording~\cite{park2022review}. 
Accurate speaker diarizaiton is crucial for a wide range of applications, such as live captioning, multi-speaker automatic speech recognition (ASR), and conversational analytics. 
Recent advances in neural network-based diarization have led to significant performance gains, driven by the availability of large public datasets and increased model capacity. 
Notably, end-to-end neural diarization (EEND) frameworks~\cite{horiguchi2022encoder} and target speaker voice activity detection (TS-VAD) systems~\cite{medennikov2020target} have gained attention for their strong empirical results. 
At the same time, traditional modular approaches to speaker diarization have also seen substantial improvements~\cite{dawalatabad2021ecapa, klement2024discriminative, raghav2025self}.

\begin{figure}
    \centering
    \includegraphics[width=1\linewidth]{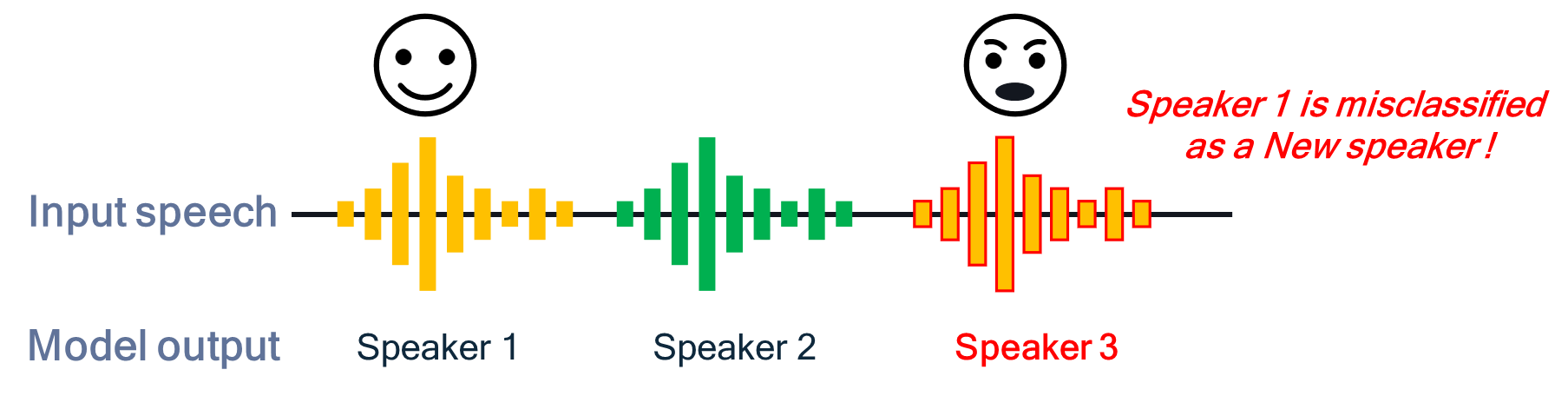}
    \caption{Intrinsic speaker variability can confuse speaker diarization systems. The system classifies the angry voice of \textit{speaker 1} as a different speaker (\textit{speaker 3}).}
    \label{fig:fig1}
\end{figure}

Despite recent progress, existing speaker diarization systems remain vulnerable to intra-speaker variability, which significantly degrades accuracy. 
This variability broadly falls into two categories: extrinsic and intrinsic.
\textit{Extrinsic variability}, caused by background noise, room impulse response, and recording devices, can often be reduced with data augmentation~\cite{barker2017chime, ko2017study}.
In contrast, \textit{intrinsic variability} originates from speaker-specific factors, such as emotional state, health conditions, or aging, which we collectively refer to as speaking style. 
As illustrated in Figure~\ref{fig:fig1}, such variability often leads to misclassification, where segments from the same speaker are mistakenly assigned to different identities.
Unlike extrinsic variability, intrinsic variability cannot be easily mitigated through data augmentation because large-scale datasets covering diverse speaking styles per speaker are lacking.

Meanwhile, recent advances in speech generation have introduced style-controllable speech synthesis~\cite{10095404, yao2025stablevc}.
Global style tokens (GST)~\cite{wang2018style} demonstrated an unsupervised approach to capture latent speaking styles using attention-based modules, and Vevo~\cite{zhang2025vevo} further proposed a framework that conditions on both text and reference speech to provide fine-grained control over stylistic variations such as emotion and expressiveness.

\begin{figure*}[t]
    \centering
    \includegraphics[width=0.88\linewidth]{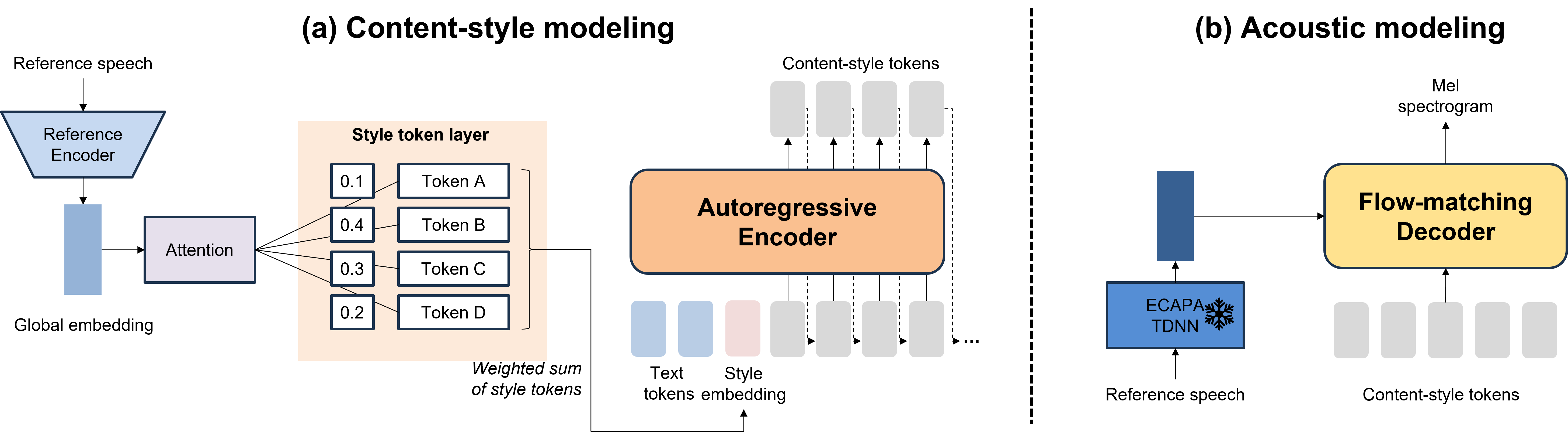}
    \caption{Data augmentation framework. An autoregressive encoder generates content-style tokens from text tokens and a style embedding. A flow-matching decoder then generates mel-spectrograms conditioned on the speaker embedding and the generated tokens from a previous stage.}
    \label{fig:fig2}
\end{figure*}

\begin{figure}[t]
    \includegraphics[width=1\linewidth]{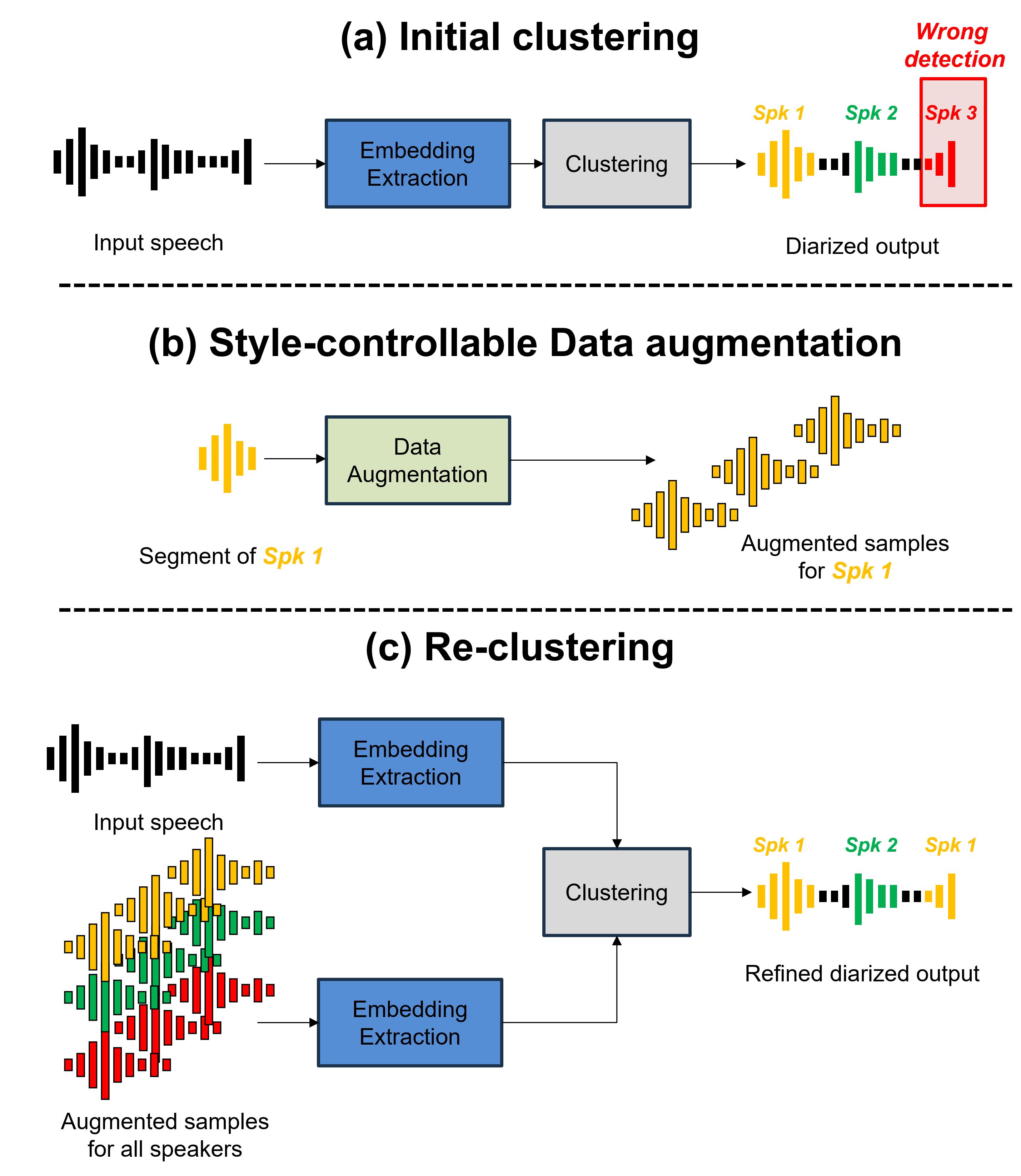}
    \caption{Schematic of the proposed framework. (a) The first stage produces initial diarized segments per speaker. (b) In the second stage, these segments are used to generate style-augmented samples. (c) Finally, the original and generated samples are combined to enhance clustering robustness by enriching stylistic diversity.}
    \label{fig:fig3}
\end{figure}

Motivated by these developments, we propose a framework that compensates for intrinsic speaker variability by leveraging style-controllable speech synthesis for data augmentation.
Our model integrates the strengths of GST and Vevo, learning style representations in an unsupervised manner while enabling flexible style control at inference.
This design makes it possible to generate unlimited stylistically diverse speech, all while preserving the identity of the original speaker.
The generation system starts with diarized segments obtained from an initial diarization stage. 
For each diarized segment, it generates augmented speech samples enriched with phonetic and stylistic diversity. 
These augmented segments are then re-embedded, and the embeddings from both the original and generated audio are blended to reinforce clustering robustness. 
This process enhances the system’s ability to group segments with high intra-speaker variability, mitigating splitting errors in scenarios with an unknown number of speakers.
Although this work was mainly designed based on a modular speaker diarization system, the proposed method can be integrated into any diarization pipeline, without requiring retraining the core model.

\iffalse
The contribution can be summarized as follows;
\begin{itemize}
    \item \textbf{Intrinsic variability compensation:} 
    Our generative augmentation framework produces style-diverse speech samples that preserve speaker identity, and the diarization framework leverages these to derive enriched embeddings to obtain robust clustering against intra-speaker variability.

    \item \textbf{Plug-and-play modularity:} 
    Our data augmentation system can be seamlessly integrated into existing speaker diarization pipelines without retraining the entire system. 
    This plug-and-play property enables flexible augmentation of style diversity across different diarization backbones.
\end{itemize}
\fi

%% file: contents/2_proposed.tex
\section{Methodology}
Our goal is to address a speaker splitting issue caused by intrinsic speaker variability, using augmented samples generated by a style-controllable text-to-speech (TTS)–based data augmentation model. 
%In Sections~\ref{subsec:method_1} and ~\ref{subsec:method_2}, we describe the implementation of each stage of the augmentation framework. 
%Then, in Section~\ref{subsec:method_3}, we explain how the augmentation model was integrated into the speaker diarization pipeline.

\subsection{Data augmentation model}
\label{subsec:data_augmentation}
Inspired by Vevo~\cite{zhang2025vevo}, we designed a two-stage augmentation framework that enables effective control over diverse speaking styles. 
Figure~\ref{fig:fig2} illustrates the framework consisting of two stages.
The first stage focuses on style manipulation, while the second stage ensures conditioning on target speaker’s identity.

\noindent\textbf{Content-style modeling}: The goal in this stage is to estimate content-style tokens conditioned on input text transcriptions and style embedding. 
The content-style tokens can be regarded as latent embeddings that jointly encode linguistic content and prosodic style, disentangled from speaker identity.
In Vevo, such disentanglement was achieved using a variational autoencoder (VAE) applied to HuBERT features with a limited codebook size.
For further details on content–style representations, we refer readers to Vevo~\cite{zhang2025vevo}.

In contrast to Vevo, which requires reference speech to guide the preferred speaking style, our framework leverages Global Style Tokens (GST)~\cite{wang2018style} to learn speaking styles, such as emotion, speed, and animation, without any explicit labels with an autoregressive transformer. 
Specifically, the reference encoder extracts global embedding from mel-spectrogram features. 
The style token layer consists of 10 trainable tokens, and multi-head attention computes the similarity between global embedding (query) and each token (key \& value). 
The resulting attention weights are used to generate a weighted sum of tokens, forming the final style embedding. 
This style embedding is concatenated with text token embeddings derived from the input transcription, and the model is trained using a next-token prediction objective in a self-supervised manner.
During the augmentation phase, a diverse amount of stylized speech could be obtained by modifying the attention weights. 

\noindent\textbf{Acoustic modeling}:
The goal in this stage is to generate acoustic features (mel-spectrogram) conditioned on the content-style token obtained from previous stage and target identity from reference speech.
Unlike Vevo, which relies on in-filling tasks to train the decoder, resulting in speaking style leakage, our model uses ECAPA-TDNN speaker embeddings using reference speech to provide speaker identity to the model. 
We design a conditional-flow matching transformer that has already proved its powerful generation performance~\cite{mehta2024matcha}. 
To extract raw waveform from a mel-spectrogram, we rely on a pre-trained BigVGAN vocoder~\cite{lee2022bigvgan}.

\subsection{Integrating data augmentation into speaker diarization pipeline}
\label{subsec:speaker_diarization}
Figure~\ref{fig:fig3} illustrates our three-stage pipeline, consisting of initial clustering, style-controllable data augmentation, and re-clustering. 
Each stage is executed sequentially.

\noindent\textbf{Initial clustering}: To conduct data augmentation, speaker-specific samples are required. 
However, only speech mixed with multiple speakers is available at the initial stage. 
To extract speaker-specific segments from the input recording, we first perform clustering for speaker embeddings of recording. 
Speaker embeddings are extracted using a pre-trained ECAPA-TDNN encoder with 1 second window and 0.2 second hop lengths. 
Spectral clustering~\cite{von2007tutorial} is then applied to these embeddings using a similarity threshold of 0.15 to obtain initial speaker groupings.

\noindent\textbf{Style-controllable Data augmentation}: Using the speaker-specific segments obtained from the initial clustering, we generate augmented samples via a style-controllable data augmentation model. 
ECAPA-TDNN speaker embeddings are compared using cosine similarity, and only pairs with a similarity score of 0.4 or higher are considered to belong to the same speaker. 
To diversify speaking styles, we manipulate GST attention weights during synthesis.

\noindent\textbf{Re-clustering}: We extract speaker embeddings from both the augmented samples and the original speech. 
These embeddings are then combined and subjected to a second round of spectral clustering using a similarity threshold of 0.12. 
To ensure balanced representation, we balance the number of speaker embeddings per speaker across original and augmented sources to avoid bias during clustering.

%% file: contents/3_exp_setup.tex
\section{Experimental setup}
\subsection{Data}
The LibriTTS-R~\cite{koizumi2023libritts} multi-speaker dataset was used to train both the autoregressive encoder and the conditional flow-matching decoder.
Specifically, the 360-hour train-clean and dev-clean subsets are employed for training and validation, respectively. 
For evaluation, two datasets were used: Concatenated emotional speech and Truncated AMI corpus.

\noindent\textbf{Concatenated emotional corpus}: To assess the impact of style variability, we concatenated samples from a real emotional speech corpus to reflect conversation scenarios with high speaking diversity.
From the ESD dataset~\cite{zhou2022emotional}, samples from 10 English speakers were taken, which include five emotions (neutral, sad, happy, surprised, and excited) per speaker.
Each sample contains two speakers with turn-taking patterns between speakers. 
The duration of each sample ranges from 30 seconds to 1 minute, and 100 samples were created.

\noindent\textbf{Truncated AMI corpus}: To evaluate model performance on real conversational data, the MixHeadset subset of the AMI corpus~\cite{kraaij2005ami} was used. 
Note that AMI recordings are typically long, ranging from 30 minutes to 1 hour, which can dilute the impact of intra-speaker variability and complicate analysis. 
Preliminary experiments showed that diarization systems are more sensitive to intra-speaker variability when recordings are short, as a limited number of data is available for clustering.
To highlight this impact, we truncated the AMI recordings into shorter segments~\cite{parthasarathy2017study}. 
Specifically, we randomly extracted audio segments of 15, 30, 60, 120, and 240 seconds, retaining only those containing exactly three speakers. 
This process was repeated until 100 samples were collected for each duration.

\subsection{Implementation details}
We trained an autoregressive Transformer encoder for 550k steps with a batch size of 4, and a conditional flow matching model for 1M steps using the same batch size. 
Both models were optimized using the Adam optimizer~\cite{kingma2015adam} with a learning rate of 2e-4. 
For acoustic feature extraction, we utilized 128-dimensional mel-spectrograms with an 80 ms window size and a 20 ms hop length, respectively.
To evaluate model performance, we synthesized augmented samples using 32 diffusion steps. 
Ground-truth voice activity detection (VAD) was applied prior to spectral clustering on speaker embeddings, and speech overlap regions were excluded from evaluation to ensure consistency. 
We assessed diarization performance using standard metrics with a speaker diarization scoring toolkit ~\footnote{https://github.com/nryant/dscore}: diarization error rate (DER), false alarm (FA), miss rate (Miss), and confusion (Conf), where lower values indicate better performance across all metrics. 
Additionally, we examined the estimated number of speakers derived from the clustering output to evaluate speaker count accuracy.

%% file: contents/4_exp.tex
\section{Experimental results}

\begin{table}[t]
\centering
\caption{DER and its breakdown in percent on Concatenated emotional corpus. The estimated number of speakers (Nspk) is also provided.}
\vspace{10pt}
\label{tab:tab1}
\begin{tabular}{c|rrrr|r}
\hline
Augmentation & DER  & Miss & FA   & Conf  & Nspk\\ \hline \hline 
X            & 10.71& 0.00 & 0.00 & 10.70 & 3.06\\
O            & 5.48 & 0.00 & 0.00 & 5.48  & 2.76\\ \hline 
\end{tabular}
\end{table}

\begin{figure}[htbp]
    \centering
    \begin{subfigure}[b]{0.48\columnwidth}
        \includegraphics[width=1\linewidth]{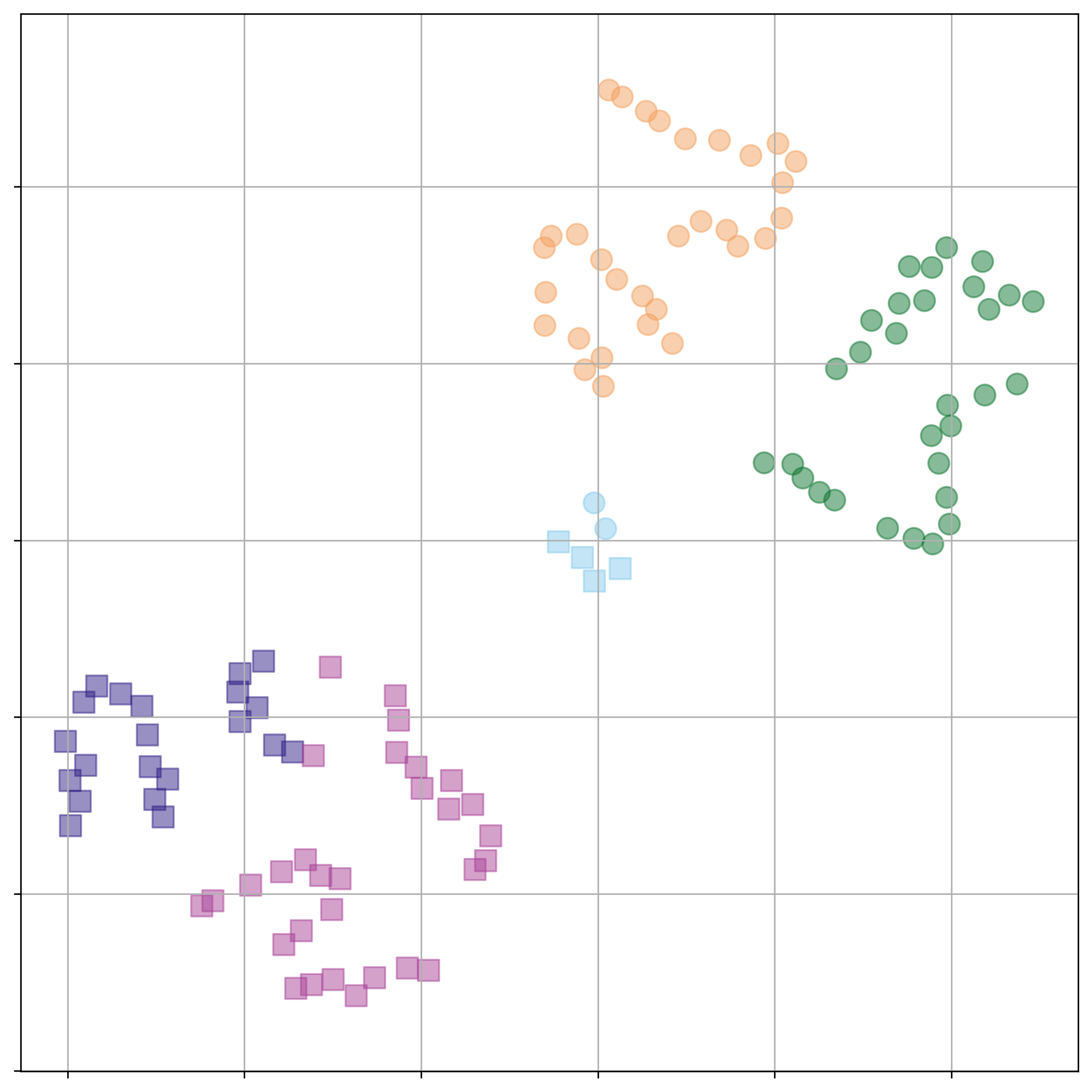}
        \caption{No augmentation}
        \label{fig:fig4a}
    \end{subfigure}
    \begin{subfigure}[b]{0.48\columnwidth}
        \includegraphics[width=1\linewidth]{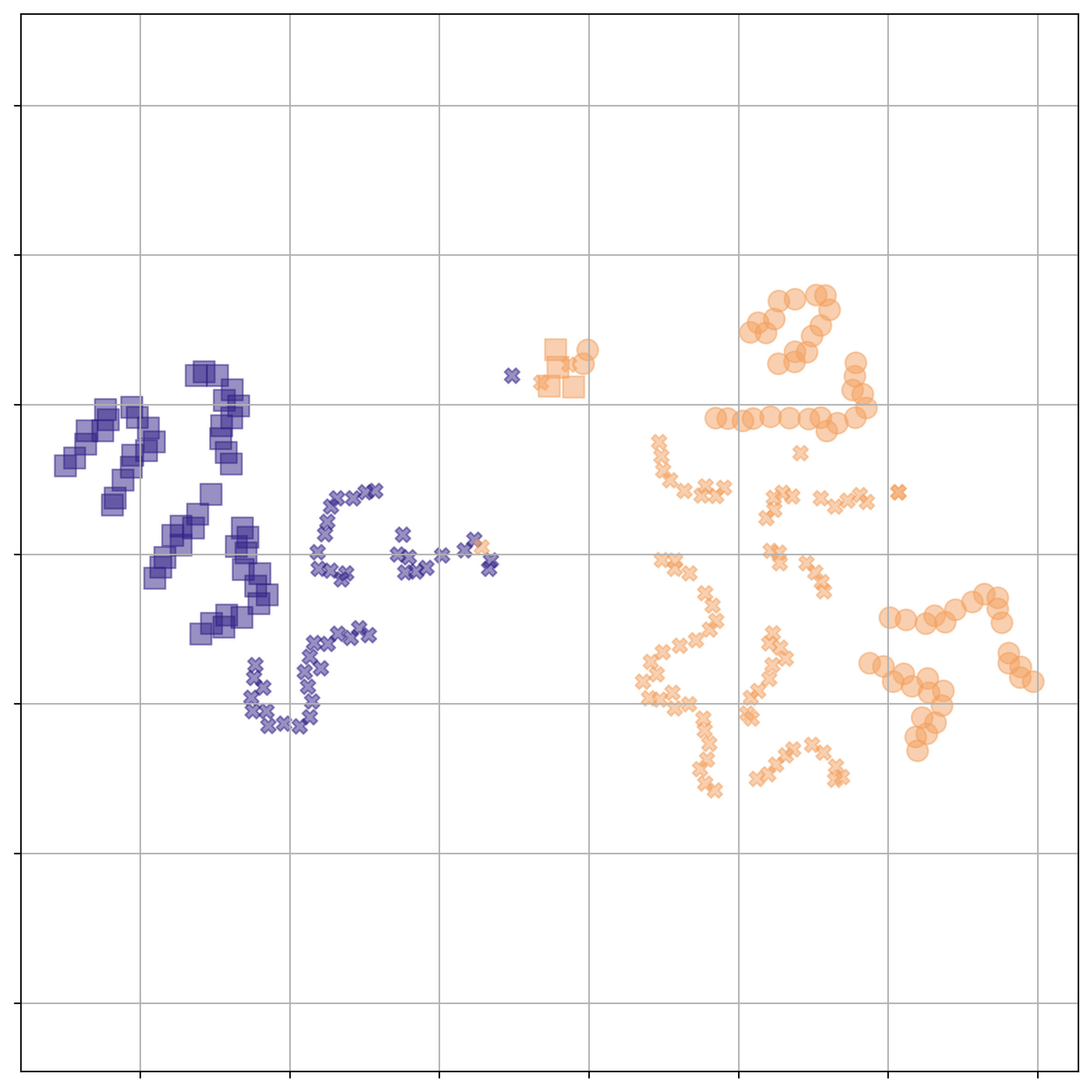}
        \caption{With augmentation}
        \label{fig:fig4b}
    \end{subfigure}
    \hfill
    \caption{t-SNE visualization of speaker embeddings. Colors indicate system-assigned speaker labels; marker shapes reflect ground-truth labels. Cross-shaped markers denote augmented samples. After re-clustering by blending embeddings from augmented and original speech, five initial clusters merge into two coherent speaker groups.}
    \label{fig:fig4}
\end{figure}
\vspace{-20pt}

\subsection{Performance on Concatenated emotional corpus}
% (miseul): When preparing the rebuttal, you should consider overestimating the number of speakers.
The DER and its breakdown calculated on the Concatenated emotional corpus are presented in Table~\ref{tab:tab1}. 
Miss and FA are zero due to the use of the oracle VAD labels. 
Data augmentation significantly reduces the confusion rate, primarily by mitigating speaker splitting errors.
Without augmentation, the system tends to overestimate the number of speakers.
In contrast, with augmentation, the estimated number of speakers aligns more closely with the ground-truth speaker count, achieving a 49\% decrease in error rate.
Figure~\ref{fig:fig4} shows the t-SNE visualization of speaker embedding to illustrate the effect of data augmentation.
After re-clustering with blended embeddings from augmented (cross-shaped markers) and original speech, five initial clusters are consolidated into two coherent speaker clusters.
Specifically, in (a), one speaker (square markers) is split into two clusters (violet and pink), another (round markers) into two others (yellow and green), with an additional mixed cluster (light blue). 
In (b), augmentation helps merge square markers into a single cluster (violet), and round markers into another (yellow).
The augmented samples effectively `bridge' the initially separated clusters by enriching style diversity and improving same-speaker grouping.

\subsection{Performance on Truncated AMI corpus}
\begin{figure}
    \centering
    \includegraphics[width=0.85\linewidth]{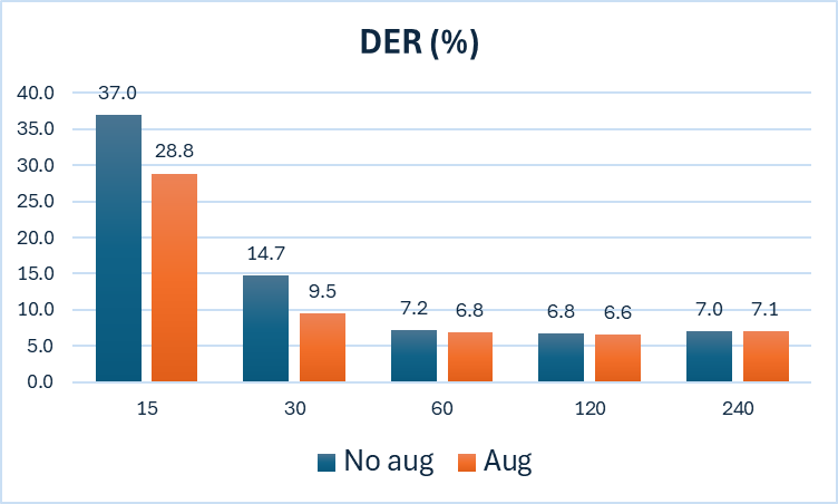}
    \caption{Performance depending on varying recording lengths for Truncated AMI corpus. Applying data augmentation reduces DER for short utterances (15- or 30-second).}
    \label{fig:fig5}
\end{figure}

System performance on real conversations using the truncated AMI corpus is shown in Figure~\ref{fig:fig5}.
DER breakdown was omitted because the use of oracle VAD label eliminated Miss and FA errors, leaving DER to directly reflect only the confusion error rate.
Without augmentation, the system maintained relatively stable performance on long excerpts (60 seconds or longer), but its performance degraded dramatically on short segments (e.g., 15- or 30-seconds).
This degradation occurs because fewer speaker embeddings are available, making it difficult to form reliable clusters. 

Applying data augmentation could alleviate the performance degradation by 22\% on 15-second segments and 35\% on 30-second segments, demonstrating that augmentation effectively compensates for the limited number of embeddings.
Although segments longer than 60 seconds showed minimal performance improvement due to already stable clustering, it is noteworthy that augmentation did not adversely impact the system performance.
%It should be noted that the current system relies on the oracle VAD labels. 
%This reliance eliminates Miss and FA errors, leaving DER to directly reflect only the confusion error rate. 
%In future work, we plan to remove this dependency and evaluate performance under more realistic VAD conditions

\iffalse
\begin{table}[t]
\centering
\caption{Performance on Truncated AMI corpus. We compare diarization results for recordings of 30 seconds and 60 seconds}
\vspace{10pt}

\label{tab:tab2}
\begin{tabular}{c|c|ccccc}
\hline
\multicolumn{1}{l|}{Duration(s)} & Aug & DER  & Miss & FA   & Conf & Nspk \\ \hline\hline
\multirow{2}{*}{30}              & X   &14.73 & 0.18 & 0.00 & 14.54& 4.07 \\
                                 & O   &9.54  & 0.18 & 0.00 & 9.36 & 3.31 \\ \hline
\multirow{2}{*}{60}              & X   &7.18  & 0.08 & 0.00 & 7.09 & 3.18 \\
                                 & O   &6.82  & 0.08 & 0.00 & 6.73 & 3.28 \\ \hline
\end{tabular}
\end{table}
\fi

%% file: contents/5_conclusion.tex
\section{Conclusion}

We addressed the speaker diarization degradation caused by intrinsic intra-speaker variability through data augmentation. 
A two-stage framework was introduced, incorporating a TTS-based style-controllable speech generation model to produce diverse speech styles, while preserving speaker identity. 
Incorporating augmented samples into clustering reduced speaker splitting in both the concatenated emotional speech and real conversation excerpts. 
Future work includes integrating a VAD module to remove oracle dependence and extending the framework to end-to-end systems.